\documentclass[pre,twocolumn]{revtex4}
\usepackage{amsmath,amscd,amsfonts,amssymb,color,hyperref}
\usepackage{graphicx,amsfonts,epsf}
\usepackage{float}

\begin{document}


\title{MULTI-LEVEL QUANTUM DIESEL ENGINE OF NON-INTERACTING FERMIONS  IN A ONE DIMENSIONAL BOX
 } 



\author{Satnam Singh}
\email{satnamphysics@gmail.com,\\satnamsingh@iisermohali.ac.in}
\affiliation{Department of Physics, Rayat Bahra University, V.P.O. Sahauran,  Distt. Mohali, Kharar, Punjab 140104}


\author{Shishram Rebari}
\email{rebaris@nitj.ac.in}
\affiliation{Department of Physics, Dr. B.R. Ambedkar National Institute of Technology Jalandhar, Punjab- 144011}


\date{\today}

\begin{abstract}
We consider the toy model of quantum Diesel cycle without temperature constructed from non-interacting fermions, which are trapped in a one-dimensional box. 
The work and energy are extracted from the cycle are by changing the expectation value of Hamiltonian. 
We analytically calculated the efficiency of the cycle and efficiency at maximum work as a function of compression ratio. We found that the efficiency of the engine depends on both compression ratio and cut-off ratio. In contrast, the efficiency at the maximum work can be written as a function of the compression ratio only.  Moreover, we calculate the Clausius relation of the cycle.  The degree of the irreversibility of the cycle depends only on the cut-off ratio.
We also study the relation between power and efficiency of the cycle.
 The power output is also studied as the function of the compression ratio. It is found that for a given value of the cutoff ratio, the dimensionless power output decreases as the compression ratio increases.

\end{abstract}

\pacs{03.67.Lx, 03.67.Bg}

\maketitle 

\section{INTRODUCTION}
 A heat engine is a well-known example of thermal machine\cite{spin-quantum-SA,zhang2014quantum,seah2018work,singh2018low}. It absorbs the heat energy and converts it into the mechanical work\cite{mani2018graphene,abah2014efficiency}.
 These thermal machines certainly play an essential role in our daily
life. These cars, air-conditioners,  lasers, refrigerators, and power plants are all examples of the thermal machines\cite{rezek2006irreversible}.
Nowadays, with the developments of quantum information processing and  nanotechnology, the study of the interface between thermodynamics and quantum physics is attracting the attention of the physicists and as well as engineers\cite{agarwalla2017quantum,abah2019shortcut,kieu2004second,thomas2012informative,iyyappan2019efficiency}.

 In the early 19th century, Sadi Carnot proposed an abstract model of a classical heat engine which was reversible and cyclic\cite{sutantyo2015quantum,wang2018nonlinear,goswami2013thermodynamics,singh2019three,singh2017feynman}. The efficiency of the Carnot engine was given by, $\eta_c=1-T_C/T_H$. The efficiency of that engine was independent of the working substance of the engine. It was depended only on the ratio of the temperature of the cold($T_C$) and the hot bath($T_H$). 
 Practically all heat engines operate far from the ideal maximum efficiency limit set by Carnot\cite{rezek2006irreversible}.
 In the modern age, technologies can miniaturize things down to the nanoscale level, where quantum effects are not negligible\cite{watanabe2017quantum,gupta2017stochastic}.
So the classical thermodynamic theory based on systems of macroscopic size does not applicable anymore on the quantum mechanical systems, and the quantum-mechanical generalization of thermodynamics becomes necessary\cite{accikkalp2018performance,narasimhachar2015low,gour2015resource}.
Due to this, the applicability of thermodynamics in the quantum regime becomes more interesting.
So it is essential to study these quantum systems directly in relation to the thermodynamic systems. There are various studies available in literature which describes the quantum version of the thermodynamical processes and cycles\cite{wang2012optimization,latifah2014multiple,latifah2013quantum}.
The central concern of all these studies is to study the quantum mechanical version of classical thermodynamic cycles and  processes\cite{lu2019optimal,gardas2015thermodynamic}.

In the classical heat engine, classical material is used as working substance, while  in the quantum heat engine, the quantum matter is used as the working substance\cite{yin2017optimal,bender2000quantum,chand2017single,dorfman2018efficiency,humphrey2005quantum,zhang2014quantum}. 
That quantum material includes the spin systems, non-interacting harmonic oscillators, particles in box systems\cite{chand2018critical,chand2017measurement,barontini2018ultra,deffner2010quantum,agarwal2013quantum,dattagupta2017ericsson,turkpencce2017photonic,uzdin2016quantum,alicki2015non,ramezani2019impact,humphrey2002reversible}.
 Scovil and Schultz-Dubois first introduced the concept of quantum thermal machines by showing the equivalence of three-level maser to the Carnot heat engine\cite{scovil1959three}.

In  literature, various examples of quantum thermodynamical cycles can be found\cite{watanabe2017quantum,deffner2016quantum,uzdin2015equivalence}. These examples include the quantum analogues of Carnot, Stirling, Otto or Diesel cycles\cite{wang2013efficiency,latifah2013quantum,sutantyo2015quantum,thomas2017implications}.
 The basic conceptual difference between these classical and quantum cycles is that in the quantum cycle one  deals with the discrete energy levels of the system\cite{single-ion-obina,zhang2014quantum,harbola2012quantum,correa2014optimal}.

In 2000, Bender et al. gave a mathematical formulation of the Carnot  engine consisting of a single quantum particle, confined in a one dimensional square well potential.  In that work, they replaced the heat baths with the energy baths and the energy eigenvalues were calculated from the Schrodinger equation\cite{bender2000quantum,latifah2012quantum,schrodinger1989statistical}.
The cycle consists of adiabatic and isoenergetic  quantum processes which are analogues of the corresponding classical Carnot cycle. After this lot of papers related to the particle in box quantum  engines have been published in this area\cite{wang2012optimization,latifah2014multiple,latifah2013quantum,bender2002entropy,sutantyo2015quantum,yin2017optimal,chattopadhyay2019relativistic}.

The purpose of the present work is to study the efficiency of a quantum Diesel cycle constructed from non-interacting fermions trapped in a box.
In this paper, we calculate the efficiency of the cycle, efficiency at maximum work, Clausius relation as the function of the ratios of the lengths. Moreover, we studied the relation of dimensionless power and efficiency of the cycle.


  \begin{figure}
 \centering
 \includegraphics[scale=1,keepaspectratio=true]{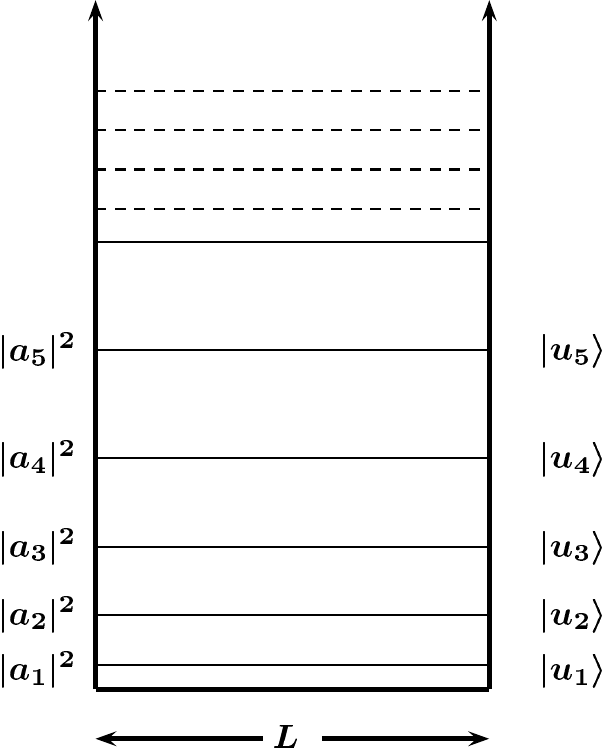}
 \caption{Pictorial representation of the levels of the box  and the probabilities of the corresponding levels} \label{fig:box}
\end{figure}
 
\section{ THERMODYNAMICS  OF PARTICLE IN BOX}
 Initially, we consider that there is a particle trapped in one-dimensional box of length $L$ having mass $m$.  The energy eigenvalue corresponding to $n^{th}$ level is given as

\begin{equation}
 \epsilon_n=\frac{\pi^2\hbar^2 n^2}{2mL^2}\label{eq:energyi}
\end{equation}
 Consider that the $|\psi\rangle$ is the wave function and it is spanned by as $|\psi\rangle= \sum_{n=1}^{\infty} a_n | u_n \rangle$(see figure \ref{fig:box}).
 The probability of the $n^{th}$ level is  $p_n=|a_n|^2$ and  $\sum_{n=0}^{\infty}|a_n|^2=1$. 
The average energy of the system is written as
\begin{equation}
 U=\sum_n\epsilon_n p_n \label{eq:internal-energy}
\end{equation}

Where $U$ analogues to the internal energy of the system. Differentiating both sides, we will get 

\begin{equation}
 dU=\sum_n(p_nd\epsilon_n +\epsilon_n dp_n) \label{eq:fstlaw-1}
\end{equation}

The first law of thermodynamics is written as
\begin{equation}
 dU=dQ-dW \label{eq:fstlaw-2}
\end{equation}
Where $Q$ is the heat and the $W$ is work done.
Comparing equation \ref{eq:fstlaw-1} with \ref{eq:fstlaw-2}, We will get

\begin{equation}
 dQ=\sum_n \epsilon_n d p_n, \hspace{0.25cm} \hspace{0.25cm} -dW=\sum_n p_n d\epsilon_n \label{eq:basic-eq}
\end{equation}
 Where $Q$ is energy and and $W$ is work. Here  both are analogues to the classical definition of heat and work.
 Now we will discuss about the Diesel cycle.

 \begin{figure}
 \centering
 \includegraphics[scale=0.9,keepaspectratio=true]{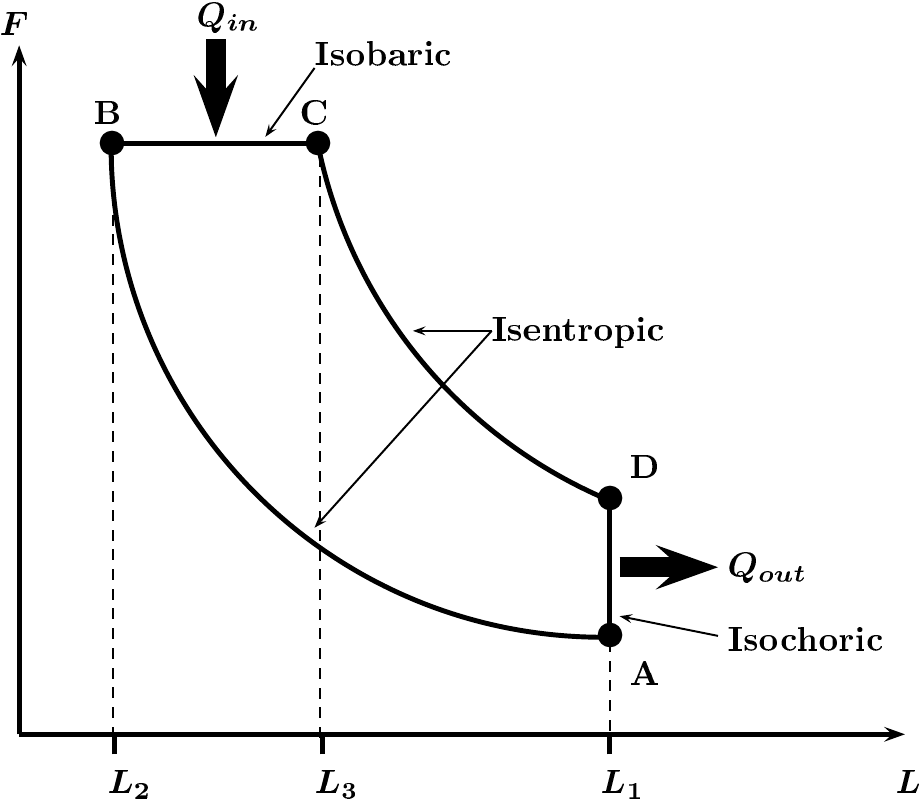}
 \caption{Plot of force and box length in Diesel cycle. $Q_{in}$ is the energy input during isobaric step and the $Q_{out}$ is the energy discarded during isochoric step.} \label{fig:cycle}
\end{figure}

\section{FERMION QUANTUM DIESEL ENGINE}
 Diesel cycle consists of the following four steps: (1) Isentropic compression, (2) Isobaric expansion (3) Isentropic expansion and (4) Isochoric compression. These steps are pictorially represented in the Figure \ref{fig:cycle}.
 In this work we consider  $N$ number of fermions of mass $m$ trapped in a one dimensional box. 
 We consider that the whole system is attached with two energy baths. One energy bath is acting as the energy source and the other energy bath is acting as the energy sink for the box.
 Due to Pauli's exclusion principle, two fermions can not occupy same state.

 We consider that the initial state of the $i^{th}$ particle of the box is $|\psi_i \rangle_{ini}$ is expanded in the terms of eigenstates $|u_n\rangle$ as $|\psi_i \rangle_{ini}=\sum_{n=1}^{\infty} a_n^{(i)}|u_n\rangle$. The total wave function is the sum of the all. The total internal energy is given as 
 
 \begin{equation}
  U=\frac{\pi^2\hbar^2 }{2mL^2}\sum_{i=1}^N \sum_{n=1}^{\infty} |a_n^{(i)}|^2 n^2  \label{eq:internal-energy-2}
 \end{equation}

 We label $\sum_{i=1}^N \sum_{n=1}^{\infty} |a_n^{(i)}|^2 n^2 = S_i$
We assume that one of the infinite potential wall  can  move infinitesimal distance. Then the energy eigenvalues  and the eigen functions vary as the function of the length of the box. 
 The force during each thermodynamical process is calculated as $F=-\partial U/\partial L$ and given by
\begin{equation}
 F=\frac{\pi^2\hbar^2 }{mL^3} \sum_{i=1}^N \sum_{n=1}^{\infty} |a_n^{(i)}|^2 n^2 
\end{equation}
Now we are able to discuss about the particle in box Diesel cycle. 

\subsubsection{Isentropic compression}

In the isentropic process, there is no energy exchange between the working substance and the external reservoir. In such a case, the  first law of thermodynamics takes the form,
\begin{equation}
 dU=-dW. \label{eq:isent-comp}
\end{equation}
The change in the internal energy is used to perform work. In the isentropic compression,  the length of the box is contracted to $L_2$ from $L_1$. Particles do not change their state.
 The work done by the system is written as 
\begin{equation}
 W_{BA}=-(E_B-E_A)=-\Big(\frac{\pi^2 \hbar^2}{2 mL_2^2}-\frac{\pi^2 \hbar^2}{2 mL_1^2}\Big) S_i
\end{equation}

 We assign  $L_1=\alpha_1 L_2$, where $\alpha_1$ is compression ratio\cite{latifah2013quantum}. Now the above equation becomes 
\begin{equation}
 W_{BA}=\frac{\pi^2 \hbar^2}{2 mL_2^2}\Big(\frac{1}{\alpha_1^2}-1\Big) S_i
\end{equation}

\subsubsection{Isobaric expansion}
Here the system is expanded in isobaric manner from $B$ to $C$.
In the isobaric expansion, the system is allowed to expand at constant pressure. 
The  first law of thermodynamics in this case is written as
\begin{equation}
 dU=dQ-F_{cons}dL
\end{equation}
In this step, the work is done by constant force and transition of the particles are also allowed.   The length of the box is expended to $L_3$ from $L_2$. The force at  both points is same. So we can write that
 \begin{equation}
  F_B=F_C,
 \end{equation}
 where $F_B$ is the force at point $B$ and $F_C$ is the force at point $C$.
energy is added in the system during this step. It is obvious that the particles will jump to the higher states. Consider that probability of the particles changes from $|a_n^{(i)}|^2$ to $|b_n^{(i)}|^2$. The force at point $C$ is given as
  \begin{equation}
 F_C=\frac{\pi^2\hbar^2 }{mL_3^3}\sum_{i=1}^N \sum_{n=1}^{\infty} |b_n^{(i)}|^2 n^2 
\end{equation}
  
  We assign  $\sum_{i=1}^N \sum_{n=1}^{\infty} |b_n^{(i)}|^2 n^2 =S_f$
 Due to this, the internal energy of the system is increased and the work is done by the system. 
 Equating the forces at point $B$ and $C$, we will get
\begin{equation}
 \frac{S_i }{L_2^3}= \frac{S_f }{L_3^3} \label{eq:ratio}
\end{equation}
 The work done by the system is given as 
\begin{equation}
 W_{CB}=-\int_B^C F dL=\frac{\pi^2 \hbar^2}{2 mL_2^2}\Big(1-\alpha_3\Big) S_i
\end{equation}
Where $\alpha_3=L_3/L_2$ is known as cut-off ratio\cite{latifah2013quantum}. The cutoff ratio is always less than compression ratio.
The  energy added from the source in the system is written as
\begin{equation}
 Q_{in}=\frac{\pi^2 \hbar^2}{2 mL_2^2}\Big(\frac{S_f}{\alpha^2_3}-3S_i+2\alpha_3 S_i\Big) \label{eq:qin}
\end{equation}
\subsubsection{Isentropic  Expansion}
From point $C$ to $D$, the system is expanding isentropically. No energy  flows from the working substance to external reservior or vice versa. The lenght of the box is expanded to $L_1$ from $L_3$. No transition occurs in this step. Work done by the system is 
\begin{equation}
 W_{DC}=\frac{\pi^2 \hbar^2}{2 mL_2^2}\Big(\frac{1}{\alpha_1^2}-\frac{1}{\alpha_3^2} \Big) S_f
\end{equation}

\subsubsection{Isochoric compression}
In the isochoric compression, the system is compressed at constant volume. In this case, the first law of the thermodynamics is written as
\begin{equation}
 dU=dQ
\end{equation}
There is no work done by the system,  but the transitions of the particles are allowed. At this stage,  the energy is removed from the system at constant box length($L_1$).  Fermions return to their initial states. The energy  removed from the system is given as
\begin{equation}
 Q_{out}=\frac{\pi^2 \hbar^2}{2 mL_1^2} (S_f-S_i) \label{eq:qout}
\end{equation}

\begin{figure}
 \centering
 \includegraphics[scale=0.6,keepaspectratio=true]{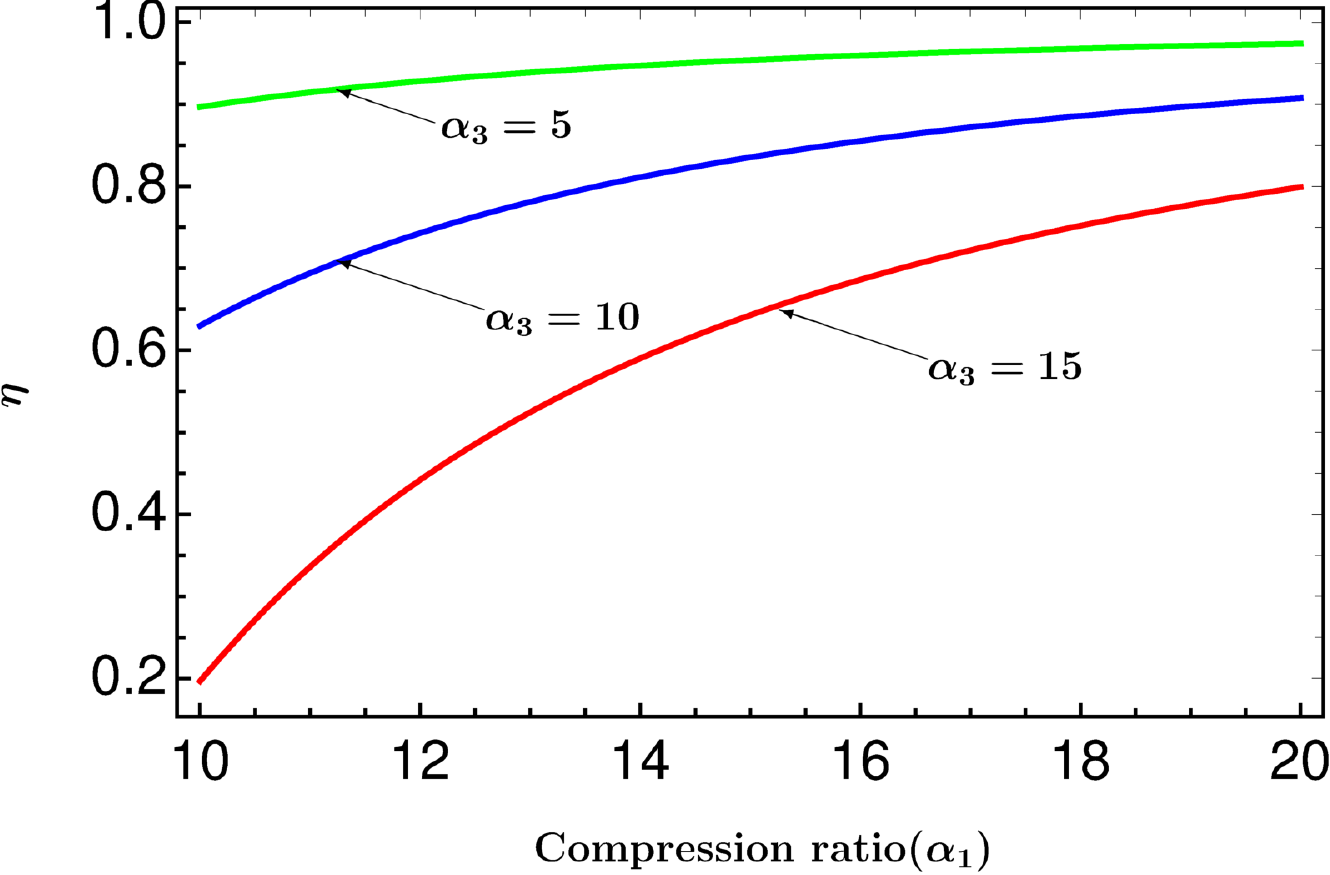}
 \caption{Efficiency of the cycle as function of compression ratio($\alpha_1$). For the particular value of cutoff ratio, efficiency of the cycle increases as compression ratio increases.} \label{fig:eff-alpha1}
\end{figure}

Total work done by the system  is the sum of the work done by the system at all stages, which is calculated as $W=W_{BA}+W_{CB}+W_{DC}+W_{AD}$ and can be written as
\begin{equation}
W=\frac{\pi^2 \hbar^2}{2 mL_2^2}\Big[\Big(\frac{1}{\alpha_1^2}-1\Big) {S_i}-2(1-\alpha_3)S_i+ \Big(\frac{1}{\alpha_3^2}-\frac{1}{\alpha_1^2}\Big)S_f\Big] \label{eq:work}
\end{equation}
Using equations (\ref{eq:work}), (\ref{eq:qin}) and (\ref{eq:ratio}),
the efficiency of the  engine is given as
\begin{equation}
 \eta=\frac{W}{Q_{in}}=1-\frac{\alpha_3^2+\alpha_3+1}{3\alpha_1^2} \label{eq:eta}
\end{equation}
Where $\eta$ is the efficiency of the particle in box Diesel engine. Same expression of efficiency was also calculated by Latifah et al for the Diesel cycle, made from a particle in one dimensional box\cite{latifah2012quantum}.  This suggest that the efficiency of the  cycle does not  depends on the number of the particles.
For the different values of the cutoff ratios, the efficiency and compression ratio is plotted in the figure (\ref{fig:eff-alpha1}).
From the figure, it is clear that for the given value of the cutoff ratio, the efficiency of the cycle increases as we increase the compression ratio.

\section{EFFICIENCY AT MAXIMUM WORK}

The total workdone by one cycle is given in equation (\ref{eq:work}). 
Differentiating it by $\alpha_3$ and equating it with zero, we get

\begin{equation}
 \alpha_3=\frac{3}{2}-\frac{1}{2 \alpha_1^2} \label{eq:efficiencies-ratio}
 \end{equation}

\begin{figure}
 \centering
 \includegraphics[scale=0.6,keepaspectratio=true]{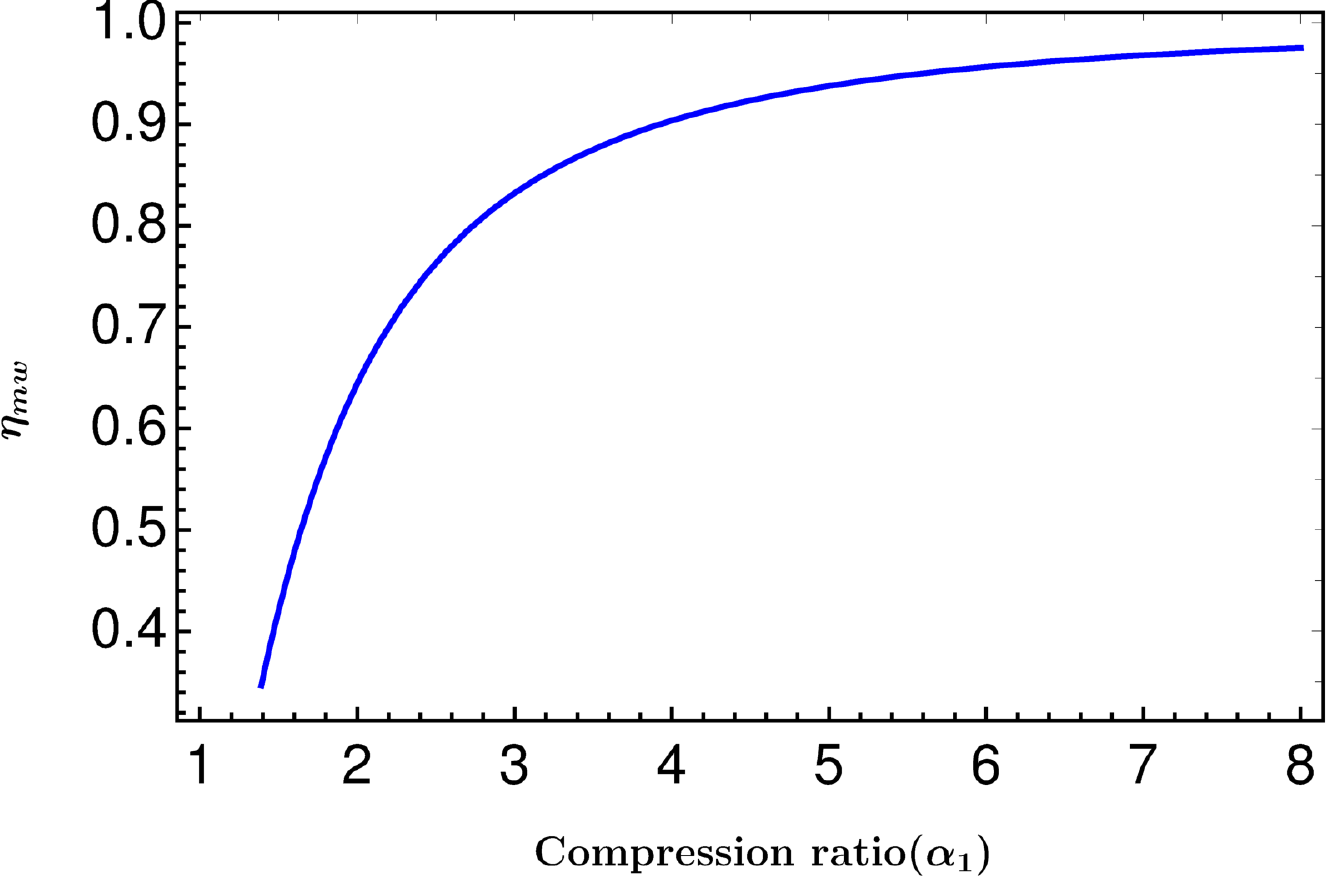}
 \caption{Efficiency at maximum work as function of compression ratio($\alpha_1$). It is clear that as we increase the compression ratio, the efficiency of the cycle also increases.} \label{fig:eff-max-cr}
\end{figure}

Substituting the value of $\alpha_3$ in equation (\ref{eq:eta}), we  get the  expression for efficiency at the maximum work$(\eta_{mw})$ as the function of compression ratio.
\begin{equation}
 \eta_{mw}=1-\frac{1-8 \alpha_1^2+19 \alpha_1^4}{12 \alpha_1^6} \label{eq:eta1}
\end{equation}
Equation (\ref{eq:eta1}) represents the efficiency at maximum work as the function of compression ratio and it  is  plotted in the Figure (\ref{fig:eff-max-cr}). From the plot it is clear that the efficiency  at maximum work of the cycle  can be increased by incresing in the compression ratio.

\section{CLAUSIUS RELATION FOR PARTICLE IN BOX  ENGINE}
Rudolf Clausius formulated an inequality to distinguish between reversible and irreversible cycles\cite{leff2018reversible}, which is known as Clausius inequality and written as
\begin{equation}
 \oint \frac{dQ}{T}\leq0
\end{equation}
where $Q$ is heat and $T$ is temperature.
 Equality holds for a reversible cycle and inequality holds for an irreversible cycle.
In case of  particle in box  cycles,  Bender et al\cite{bender2002entropy} proposed an inequality which is analogues to Clausius inequality. It is written as
\begin{equation}
 \oint \frac{dQ}{E}\leq0 \label{eq:clausis}
\end{equation}
Here $Q$ is energy and $E$ is extream limit of the energy.
Although this inequality can not used to  determine the entropy change of the system, but it gives the insight about the irreversibility of the cycle. 
Energy absorbed during the isobaric process is given by the equation (\ref{eq:qin}) and the energy dumped out in isochoric process is given by the equation (\ref{eq:qout}).
The engine is operated between two energy limits $E_H$ and $E_C$, which are given as

 \begin{equation}
  E_H=\frac{\pi^2 \hbar^2 S_f}{2 m L_3^2 }, \hspace{1cm}  E_C=\frac{\pi^2 \hbar^2 S_i}{2 m L_1^2 } \label{eq:limits}
 \end{equation}

The $ \oint {dQ}/{E}$ is written as

\begin{equation}
  \oint \frac{dQ}{E}=\frac{Q_{in}}{ E_H}+\frac{Q_{out}}{ E_C}=  4-\frac{3}{\alpha_3}-\alpha_3^3 \label{eq:inequality}
\end{equation}

\begin{figure}
 \centering
 \includegraphics[scale=0.6,keepaspectratio=true]{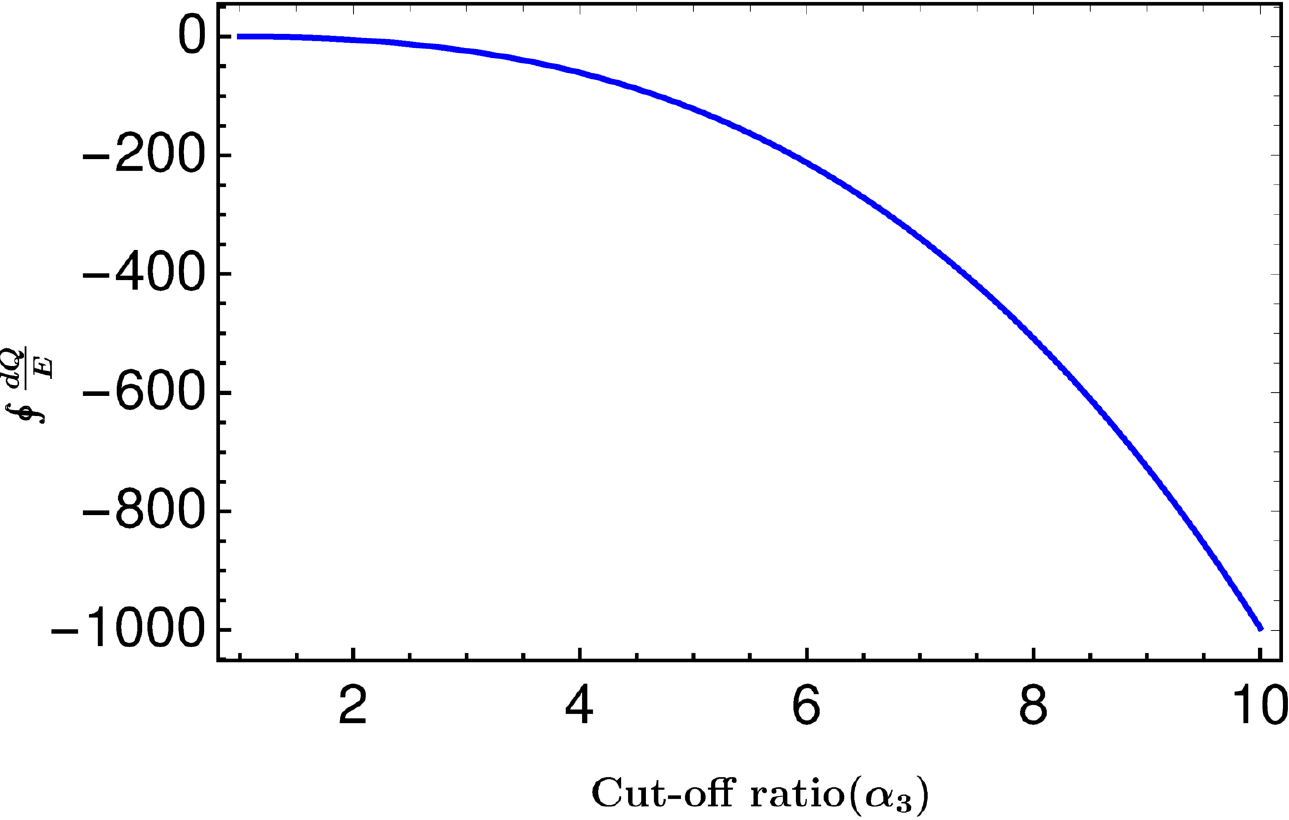}
 \caption{ $ \oint {dQ}/{E}$ as the function of cutoff ratio. As we increase the cutoff ratio, the cycle becomes more irreversible. } \label{fig:entropy}
\end{figure}

Equation  (\ref{eq:inequality}) states that the irreversibility of the cycle  does not depends on the number of particles.  It depends on the cutoff ratio.
 The inequality function and the cut-off ratio($\alpha_3$)  is plotted in Figure \ref{fig:entropy}.
 From the figure (\ref{fig:entropy}), it is clear that in Diesel cycle system holds the inequality. It is due to its irreversibility of the Diesel cycle\cite{bender2002entropy}. As we increase the cutoff ratio, the cycle become more irreversible.

 \section{OPTIMIZATION ON THE PERFORMANCE
OF THE  ENGINE}   
 
\begin{figure}
 \centering
 \includegraphics[scale=0.6,keepaspectratio=true]{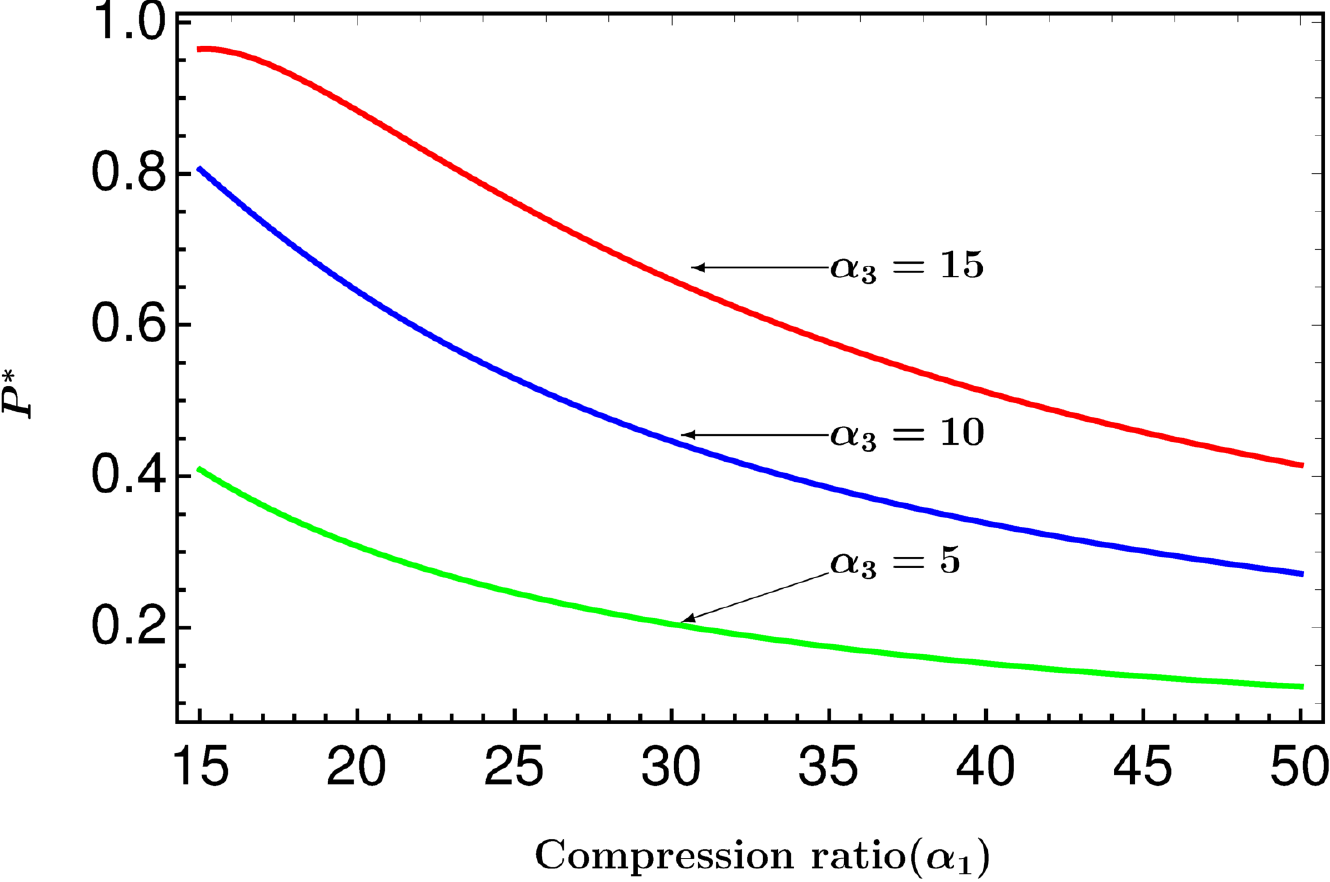}
 \caption{Dimensionless power vs compression ratio($\alpha_1$). For a given value of cutoff ratio, as we increase the compression ratio, the power of the cycle decreases} \label{fig:po-alpha}
\end{figure}

 Now we will discuss the power output and the efficiency of the Diesel cycle. The total movement of the wall is $L(=2(L_1-L_2))$,  and $\bar v$ is the average speed of the cycle and $\tau$ is the time of the one cycle. 
 In order to apply adiabatic theorem, the time scale associated with the state change should be greater than dynamical one, $\sim\hbar/E$.  In the other words, the $\bar v \ll \frac{L}{\hbar/E}$\cite{wang2012optimization,wang2012performance}.
 The total work done by the system in one cycle is given in equation (\ref{eq:work}).
The power of the system is written as
\begin{equation}
 P=\frac{W}{\tau} \label{eq:power}
\end{equation}
Where $\tau$ is time period of one cycle and it is given as
 \begin{equation}
  P=\frac{S_f \pi^2 (1+3\alpha_1^2(-1+\alpha_3)-\alpha_3^3) \hbar^2}{2 mL_2^2 \alpha_1^2 \tau}
 \end{equation}
 The cycle time($\tau$) can be written as $2(L_1-L_2)/\bar{v}$. After substituting the value of the $\tau$, the expression of the power is written as
  \begin{equation}
  P=\frac{S_f \bar{v} \pi^2 (1+3\alpha_1^2(-1+\alpha_3)-\alpha_3^3) \hbar^2}{4 mL_2^3 \alpha_1^2  (\alpha_1-1)} \label{eq:power-output}
 \end{equation}
Now we will study the relation of dimensionless power and efficiency.
 We consider that $ K=\frac{S_f \bar{v} \pi^2  \hbar^2}{4 mL_2^3 }$.
 The dimensionless power is written as
 \begin{equation}
  P^*=\frac{P}{K}=\frac{1+3\alpha_1^2(-1+\alpha_3)-\alpha_3^3 }{\alpha_1^2  (\alpha_1-1)} \label{eq:dmlpow}
 \end{equation}
The equation (\ref{eq:dmlpow}) is the relation between the dimensionless power, compression ratio and cutoff ratio.  
The Figure \ref{fig:po-alpha} displays the dimensionless power output as the function of compression ratio at different values of the cutoff ratios. From the figure it is clear that for the given value of cutoff ratio, the value of the dimensionless power decreases as the compression ratio increases. The power output also decreases when the cutoff ratio decreases.
  From equation (\ref{eq:eta}),  $\alpha_3$ is written as 
 \begin{equation}
  \alpha_3= \frac{1}{2}(-1+\sqrt{3}\sqrt{-1+4 \alpha_1^2-4 \alpha_1^2 \eta}  ) 
  \label{eq:alpha3}
 \end{equation}

 \begin{figure}
 \centering
 \includegraphics[scale=0.6,keepaspectratio=true]{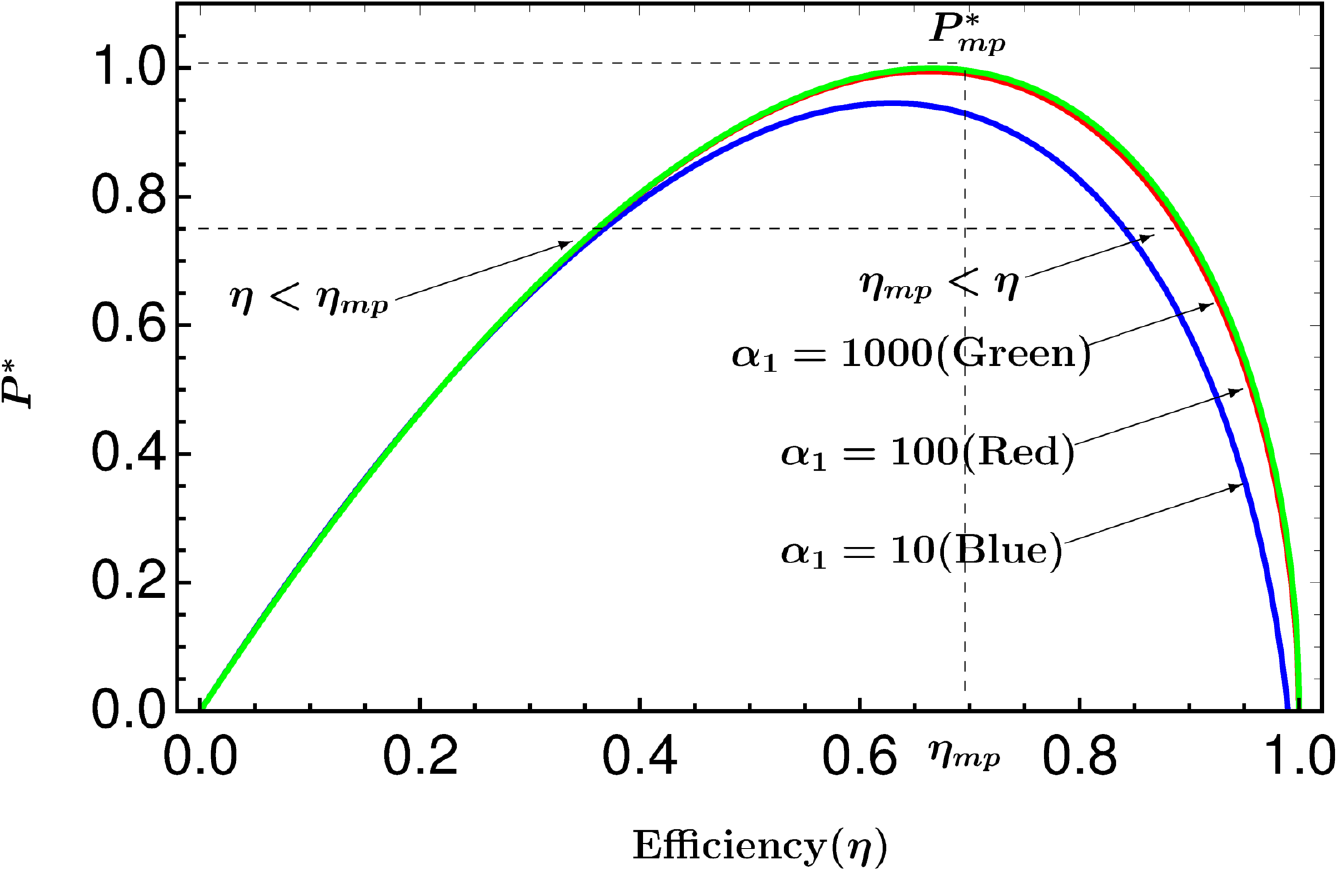}
 \caption{Dimensionless power output($P^*$) vs efficiency at the different values of the compression ratios. } \label{fig:po-ef}
\end{figure}
 
 From equations (\ref{eq:dmlpow}) and (\ref{eq:alpha3}), the dimensionless power and efficiency is related as 
 \begin{equation}
  P^*=\frac{3(-3+\sqrt{-3-12 \alpha_1^2 (-1+\eta)})\eta}{4(-1+\alpha_1)}
 \end{equation}
 For the different values of compression ratios, the characteristic curve of the dimensionless power and efficiency is given in the Figure \ref{fig:po-ef}.
It shows that curve between the dimensionless power and the efficiency is parabola like\cite{wang2012performance}.  For the larger values of compression ratio, the power output curve does not change. This show that for the larger value of compression ratio, the power of the cycle does not change.


\section{CONCLUSIONS}
In this work, we successfully studied the efficiency and power of particle in box Diesel engine constructed from the $N$ non-interacting fermions trapped in one-dimensional box. We studied the efficiency, efficiency at maximum work and Clausius relation as the function of the ratios of box lengths. Moreover, we also study the relation between efficiency and power of the cycle.
From the figures (\ref{fig:eff-alpha1}) and (\ref{fig:po-alpha}), we can conclude that for a particular value of cutoff ratio, as we increase the compression ratio, the efficiency of the cycle increases and power of the cycle decreases. The characteristic plot of the efficiency and dimensionless power is parabola like. For the maximum power of the cycle, there is one efficiency corresponding to that power. For the large value of the compression ratio, the power of the cycle does not change.
\section{ACKNOWLEDGEMENT}
Authors greatfully acknowledge Obinna Abah for the fruitful discussions.
\bibliography{engine}
\bibliographystyle{apsrev4-1}
\end{document}